\documentclass[a4paper,12pt]{article}
\usepackage{epsfig}
\usepackage{graphicx}
\usepackage{subfigure}
\begin{document}

\title{\textbf{Scattering of Topological Solitons on Barriers and Holes in Two 
$\varphi^{4}$ Models}}
\author{Jassem H. Al-Alawi\thanks{e-mail address:J.H.Al-Alawi@durham.ac.uk} and Wojtek J. Zakrzewski\thanks{email address: W.J.Zakrzewski@durham.ac.uk}
\\ Department of Mathematical Sciences,University of Durham, \\
 Durham DH1 3LE, UK\\
}

\date{\today}

\maketitle

\begin{abstract}

We present results of our studies of various scattering properties of topological solitons on obstructions in the form of holes and barriers in 1+1 dimensions. Our results are based on two models involving a $\varphi^{4}$ potential. The obstructions  are characterised by a potential parameter, $\lambda$ which has a non-zero value in a certain region of space and zero elsewhere. 
In the first model the potential parameter is included in the potential and in the second model the potential parameter is included in the  metric.
Our results are based on numerical simulations and analytical considerations.

\end{abstract}

\section{Introduction}
Some surprising aspects of the scattering of solitons have been attracting the attention of researchers and have been studied in many papers [1-3]. These studies have shown that a topological soliton
 behaves like a classical point particle when it is scattered on a potential barrier. 
In such a case the soliton follows a well defined trajectory and as it meets the barrier it slows down and, if it has enough energy to `climb' over the barrier, it gets transmitted otherwise it gets reflected with almost no loss of energy. Hence, soliton's scattering on a potential barrier
is very elastic and thus the soliton behaves like a classical point particle. However, a soliton displays more interesting behaviour when it encounters
 a potential hole. Consider a classical point particle encountering a potential hole. Then, when the particle enters the hole, the hole speeds it up,
no energy is lost, and the point particle always gets transmitted. However, at low velocities, below a critical value, the scattering of solitons on a potential hole exhibits  some aspects of a quantum like behaviour in the sense that, sometimes, the solitons get reflected by the hole although most of the time they get trapped in the hole. Of course such reflections are not possible for a classical point particle. A soliton with enough energy, {\it ie} with velocity above a critical value can get out of the hole but it comes out with much reduced velocity as during the whole process
 it radiates some of its energy [1-2].
 
In the present study, we look at various scattering properties of a topological soliton on an obstruction in two Lagrangian models. The Lagrangians are so constructed that, far away from the obstruction, the soliton in the two models looks exactly the same.

In the numerical part of this work we used square well potential barriers and holes of width 10. The simulations were performed using the 4th order Runge – Kutta method of simulating the time evolution. We used 1201 points with the lattice spacing of  $dx=0.01$. Hence, the lattice extended from -60 to 60 in the $x$-direction. The time step was chosen to be $dt=0.0025$. We used the absorbing boundary conditions.

\subsection{Model 1: $\tilde \lambda(x)\,\varphi^{4}$ potential.}

First we consider the model, in (1+1) dimensions $\mu$=0,1, defined by
\begin{equation}
\ell=\frac{1}{2}\partial_{\mu}\varphi\partial^{\mu}\varphi-U\left(\varphi\right)
\end{equation}
with the potential
\begin{center}
$U\left(\varphi\right)=\tilde\lambda\left(x\right)\left(\varphi^{2}-1\right)^{2},$
\end{center}
where $\tilde \lambda=\lambda_0+\lambda(x)$, and $\lambda(x)$ is a potential parameter which has been inserted into the Lagrangian to take into account the effects of  obstructions, holes and barriers, and so is nonzero only in a certain region of space. 

After varying the action, the equation of motion is 

\begin{equation}
\partial_{\mu}\partial^{\mu}\varphi+4\tilde\lambda\left(x\right)\varphi\left(\varphi^{2}-1\right)=0.  
\end{equation}

This equation cannot be solved analytically because the potential  has a spatial dependence.  However, this equation can be solved numerically
by placing the soliton in the region where $\lambda=0$  by making and then evolving it from there. 

Thus the potential parameter $\tilde \lambda(x)$  has a constant value, $\lambda_{0}$, in the region where the soliton is located
and far away from the position of the soliton $\tilde\lambda(x) \varphi\left(\varphi^{2}-1\right)$ is zero
as $\varphi\sim 0$.   Moreover, as the system is fully relativistic, we can get a time-dependent solution by simply boosting the static one. 
This gives us a soliton moving with velocity $u$. Thus (assuming that $x_0$ is nonzero and $t$ is small we can put

\begin{equation}
\varphi\left(x,t\right)=\pm\tanh\left(\gamma\sqrt{2\lambda_{0}}\left(x-x_{0}-ut\right)\right),
\end{equation}
where
\begin{center}
\begin{equation}
\gamma=\frac{1}{\sqrt{1-u^{2}}}.
\end{equation}
\end{center}

Here the plus and minus signs represent the kink and anti-kink solutions respectively. At $t=0.0$ the soliton is far away from the barrier/hole and this
description described the soliton very well. In what follows we put $\lambda_{0}=1.0$ and so our initial conditions for the field and its time derivative are given by

\begin{equation}
\varphi\left(x,0\right)=\pm\tanh\left(\gamma\sqrt{2}\left(x-x_{0}\right)\right)
\end{equation} 
          
\begin{equation}
\partial_{0}\varphi\left(x,0\right)=\mp\,u\gamma\sqrt{2}\, sech^{2}\left(\gamma\sqrt{2}\left(x-x_{0}\right)\right)
\end{equation}

The energy density of the field is given by

\begin{equation}
\epsilon=\frac{1}{2}\left(\partial_{0}\varphi\right)^{2}+\frac{1}{2}\left(\partial_{1}\varphi\right)^{2}+\left(\varphi^{2}-1\right)^{2}.
\end{equation}

\subsection{Model 2:  A $\lambda\varphi^{4}$ model with a position dependent metric.}
In this paper, we consider also the behaviour of  a soliton in (1+1) dimensions when  the  potential parameter  $(\lambda(x)$) arises in the space-time metric, {\it ie} the metric is given by

\begin{center}
$g^{\mu\nu}$=
\begin{math}
\left(
\begin{array}{crr}
         1+\lambda\left(x\right)&0\\
          0&-1
 \end{array}
\right).
\end{math}
\end{center}

Thus we are putting the obstruction into the model in the same way as was done in \cite{three}
in the sine-Gordon case. 

 The action is now given by

\begin{equation}
s=\int\ell\left(\varphi,\partial_{\mu}\varphi\right)\,dx\,dt,
\end{equation}

where  $\ell$ is the Lagrangian density and $g$ is the determinant of the metric, {\it ie} $g\left(x\right)^{\mu\nu}$

\begin{center}
\begin{equation}
\ell=\sqrt{-g}\left[\frac{1}{2}\partial_{\mu}\varphi\partial_{\nu}\varphi\,g^{\mu\nu}-U\left(\varphi\right)\right]
\end{equation}    
\end{center}

with the potential still being given by
 
\begin{center}
$U\left(\varphi\right)=\left(\varphi^{2}-1\right)^{2}.$
\end{center}                      
            
 Note that the coefficient of the potential in $U(\varphi)$ is chosen to be 1, {\it ie}
 it is the same as in model 1 away from the obstruction.       
The equation of motion is now: 

\begin{equation}
\left(1+\lambda\left(x\right)\right)\partial_{0}^{2}\varphi-\partial_{1}^{2}\varphi-\frac{1}{2\mid1+\lambda\left(x\right)\mid}\partial_{1}\lambda\left(x\right)\partial_{1}\varphi+4\varphi\left(\varphi^{2}-1\right)=0.
\label{a}
\end{equation}
        
The topological charge is independent of the metric and does not depend on the potential. The equation of motion (\ref{a}) cannot be solved analytically because of the spatial dependence of the potential parameter. But, like before, a solitonic solution can be found by making the potential parameter constant, $\lambda\left(x\right)=\lambda_{0}$, and then use that solution as an initial condition to solve the equation (\ref{a}) numerically. Note that the equation (\ref{a}) can be rewritten as 

\begin{equation}
\left(1+\lambda_{0}\right)\partial_{0}^{2}\varphi-\partial_{1}^{2}\varphi+4\left(1+\lambda_{0}\right)\varphi\left(\varphi^{2}-1\right)=0.
\end{equation}

And the approximately time-dependent solution for the field is given by

\begin{equation}
\varphi\left(x,t\right)=\pm\tanh\left(\sqrt{\frac{2}{1-\,u^{2}\left(1+\lambda_{0}\right)}} \left(x-x_{0}-\,u\,t\right)\right).
\end{equation}

At $t=0.0$, $\lambda_{0} =0.0$ and so the initial conditions are

\begin{equation}
\varphi\left(x,0\right)=\pm\tanh\left(\sqrt{\frac{2}{1-\,u^{2}\left(1+\lambda_{0}\right)}} \left(x-x_{0}\right)\right),
\end{equation}

\begin{equation}
\partial_{0}\varphi\left(x,0\right)=\mp\,u\sqrt{\frac{2}{1-\,u^{2}\left(1+\lambda_{0}\right)}}  \,sech^{2}\left(\sqrt{\frac{2}{1-\,u^{2}\left(1+\lambda_{0}\right)}} \left(x-x_{0}\right)\right),
\end{equation}         
which are, in fact, the same as the initial conditions of model 1.

The energy density is then given by 

\begin{equation}
\epsilon=\sqrt{\,g^{00}}\left[\frac{\,g^{00}}{2}\left(\partial_{0}\varphi\right)^{2}+\frac{1}{2}\left(\partial_{1}\varphi\right)^{2}+\left(\varphi^{2}-1\right)^{2}\right].
\end{equation}

\section{Barrier Scatterings in both models.}

The scattering of a topological soliton on a potential barrier, in model 1, was found to be nearly elastic. However, in model 2, the scattering 
was found be less elastic as, in this case, the soliton radiated away a small amount of its energy, although this radiated energy was very small  (almost negligible) when the velocity of the soliton was far below its critical value. However, it increased as we sent solitons with higher velocities.

The shape of the potential barrier, as seen by the soliton, can be found by placing the soliton with a zero velocity at different positions and calculating its total energy. This has given us figure 1 which shows  how a soliton would see a potential barrier of height 0.5 in both models 

\begin{figure}
\begin{center}
\includegraphics[angle=270, width=6cm]{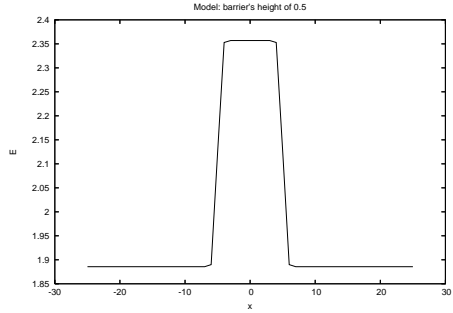}
\caption{The potential barrier as seen by the soliton in the two models}
\end{center}
\end{figure}

However, performing numerical simulations we have found a difference between the two models.
In model 1 the behaviour of a topological soliton resembles a classical point particle as only a very small amount of energy  is being radiated
out during the scattering process. However, in model 2, this amount of radiation is much larger and it increases as we increase the height of the barrier. Figure 2a and 2b shows the time evolution of the position of the soliton (moving with velocity 0.8) for different barrier’s heights in models 1 and model 2 respectively. We see very clearly that in model 2 the soliton 
loses more energy (through radiation) and consequently leave the scattrering region with lower velocity. Not surprisingly, this effect grows with
the increase in barrier's height in both models but is much more pronounced in model 2.

\begin{figure}
\begin{center}
\includegraphics[angle=270, width=6cm]{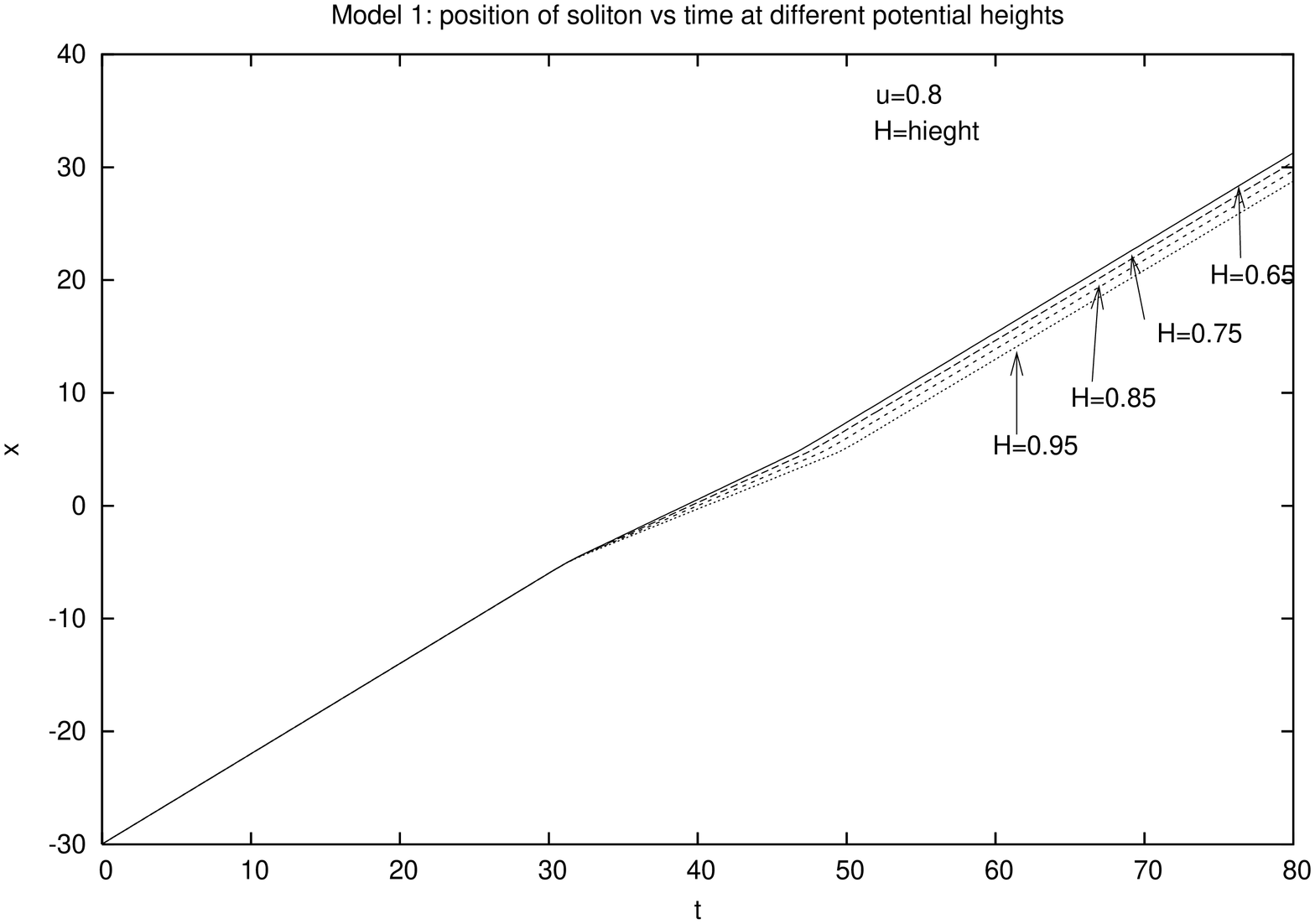}
\includegraphics[angle=270,width=6cm]{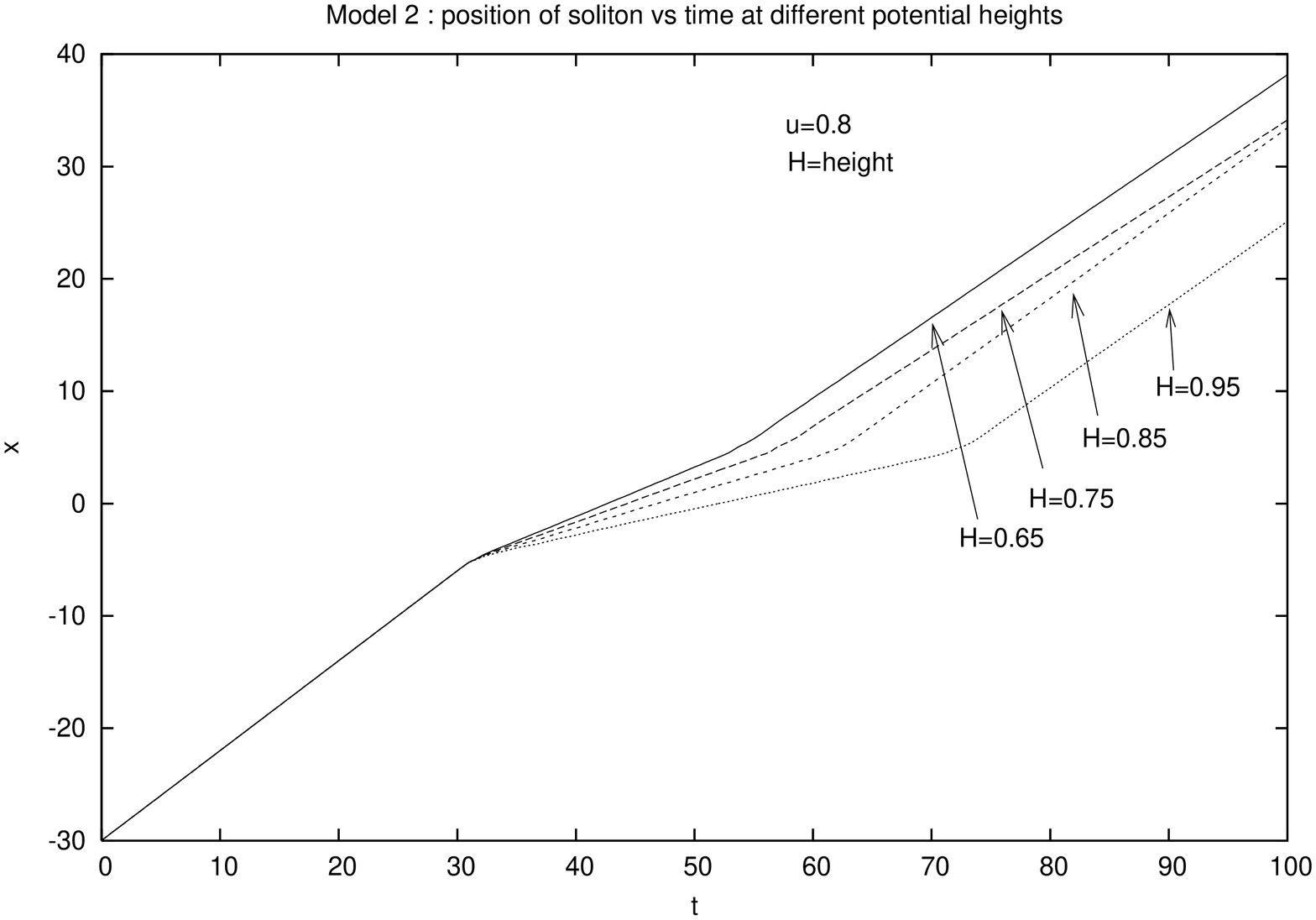}
\caption{The position of soliton vs time for various barrier's heights}
\end{center}
\end{figure}

We have also found a small variation of outgoing velocity of the soliton as a function of its incoming velocity; again this 
effect is more pronounced in model 2.  Figure 3 shows our results for a potential barrier of height = 0.5. Figure 4 indicates, how in model,  this variation Oscillation) get more pronounced with the increase the height of the barrier.

\begin{figure}
\begin{center}
\includegraphics[angle=270, width=6cm]{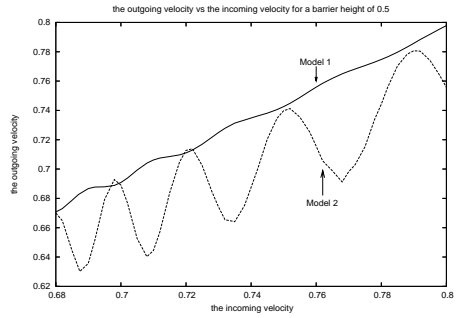}
\caption{The $v_{out}$ vs $v_{in}$ for a barrier height of 0.5 in the two models}
\end{center}
\end{figure}

\begin{figure}
\begin{center}
\includegraphics[angle=270, width=6cm]{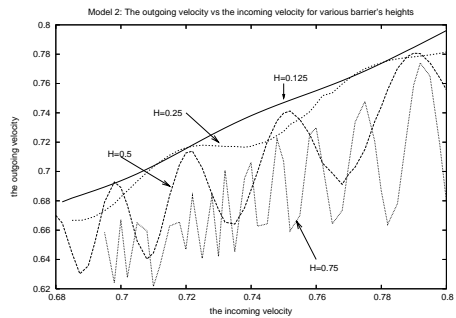}
\caption{Model 2: $v_{out}$ vs $v_{in}$ for various barrier's heights}
\end{center}
\end{figure}

\subsection{Some Analysis of Our Results.}

We have found that the solitons behave in a similar way in the two models but that the differences from the behaviour of a point particle
are more pronounced in model 2. This difference was found to be related to the oscillations of the kink corresponding to the change of its slope. These oscillations are generated because the slope of soliton changes as the soliton moves up a barrier. This encounter
is somewhat sudden, the slope changes too much and so  its starts oscillating. The soliton thus tries to readjust itself to the new value
of the slope (corresponding to the new value of $\tilde lambda$) and during this readjustment some of its kinetic energy is converted into oscillating energy. To test such an interpretation we divided the slope of the soliton before the interaction with the barrier over the slope of the soliton when it is at the barrier and then have plotted this ratio as a function of the incoming velocity. Figure 5 shows the plot of the ratio for a barrier of height =  0.5 in the two models. From the figure on can see that below the critical value, {\it ie} when the soliton gets reflected, the ratio of the two slopes is small and increases steadily as the soliton velocity increases. However, above the critical value, {\it ie} the soliton gets transmitted, the ratio in model 2 increases steadily but only slightly this time while in the model 1 it is almost constant. 

\begin{figure}
\begin{center}
\includegraphics[angle=270, width=6cm]{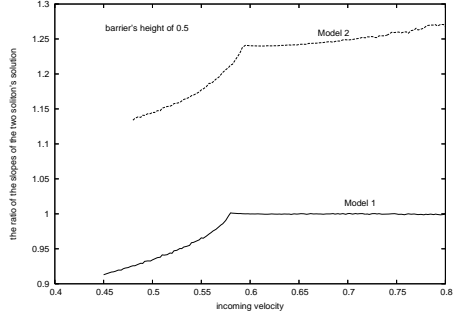}
\caption{Ratio of the slopes of soliton's solutions vs incoming velocity in the two models }
\end{center}
\end{figure}

Furthermore, we have also studied the range of the oscillations and the average square of their frequency.  We have found that above the critical
 value the average square of the frequency is directly proportional to the incoming velocity. This demonstrates that the energy radiated, which is proportional to the square of the frequency, increases as we increase the incoming velocity. This supports our argument that the soliton oscillates and that a fraction of its energy is converted into vibrating energy.

The range of the oscillation is given by
\begin{equation}
\delta=\frac{S_{1}-S_{2}}{S_{1}\sqrt{S_{2}^{2}-1}},
\end{equation}
where $S_{1}$ is the slope of the soliton before and after the barrier. $S_{2}$ is the slope at the barrier.
This result holds in the limit when the difference between the slopes is very small (so that we can make
the approximation that $\sin(\delta)\sim \delta$).
The average of the frequency, $ \omega$, is given by
\begin{equation}
\omega=\frac{\delta}{\Delta t},
\end{equation}  
where $\Delta t$ is the time spent by the soliton at the barrier. Figure 6 illustrates the relation between the average square of the
 frequency and the incoming velocity, at heights  0.5 and 0.75. The figure shows that, as the incoming velocity increases,  the oscillations 
increase and so does the radiated energy.

\begin{figure}
\begin{center}
\includegraphics[angle=270, width=6cm]{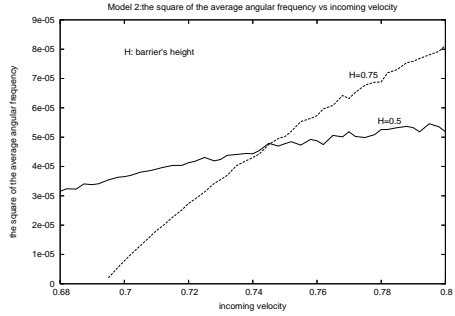}
\caption{ $\omega^{2}$ as a function of $v_{in}$ in model 2 }
\end{center}
\end{figure}

We have found that the difference in oscillations and obviously in the behaviour of solitons in the two models is due to the different 
energy, {\it ie} mass, they have at the barrier. The more massive the soliton is the fewer oscillations it makes during the interaction. There is a greater sensitivity of the potential for less massive solitons than for the heavy ones. The mass of a static soliton, $M_{rest}$, is given by the integral over the energy density:

\begin{equation}
\,M_{rest}=\int_{-\infty}^{\infty}\left[\frac{1}{2}\varphi_{x}^{2}+\,U\left(\varphi\right)\right]\,dx=\frac{8\sqrt{1+\lambda_{0}}}{3\sqrt{2}}.
\end{equation}           
        
When the soliton is far from the obstruction $\lambda_{0}=0.0$ and so its energy {\it ie} mass $M=1.88557$ in both models and this is exactly what was found numerically. For a moving soliton its is given by 

\begin{equation}
\,E=\int_{-\infty}^{\infty}\left[\frac{1}{2}\varphi_{t}^{2}+\frac{1}{2}\varphi_{x}^{2}+\,U\left(\varphi\right)\right]\,dx=\frac{M} {\sqrt{1-\,u^{2}}},
\end{equation}                                             
where $u$ is the speed of the soliton.



 Figure 7 shows a plot of the soliton rest mass as a function of the height of the barrier in the two models. 

\begin{figure}
\begin{center}
\includegraphics[angle=270, width=6cm]{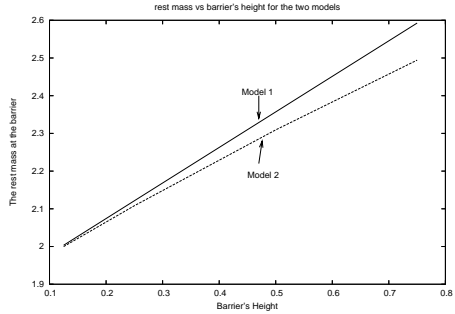}
\caption{Rest mass vs barrier's height in the two models}
\end{center}
\end{figure}

We have also looked at the values of critical velocities in both models. In the table 1
we give these values for various heights of the potential in both models. It is clear that the critical velocities in model 2 are greater than those of the model 1. That is very understandable since to reach the values of the rest masses in model 1, one needs a larger value of the gamma factor which thus leads to smaller critical velocities. 

\begin{center}
\begin{table}
\begin{tabular}{|c|}
\hline
critical velocities \\
\hline
\end{tabular}
\begin{tabular}{|c|c|c|c|c|}
\hline
Model Number & $\lambda_{0}=0.125$ & $\lambda_{0}=0.25$ & $\lambda_{0}=0.5$ & $\lambda_{0}=0.75$ \\
\hline
Model 1 & $\sim 0.34$ & $\sim0.45$ &$\sim 0.58$  & $\sim 0.66$ \\
\hline
Model 2 & $\sim 0.358$ & $\sim 0.463$ &$\sim 0.596$ & $\sim 0.695$ \\
\hline
\end{tabular}
\caption{Soliton' s critical velocities for potentials of various heights}
\end{table}
\end{center}

Figure 8 shows a plot of the critical velocity as a function of  the height barriers in the two models.

\begin{figure}
\begin{center}
\includegraphics[angle=270, width=6cm]{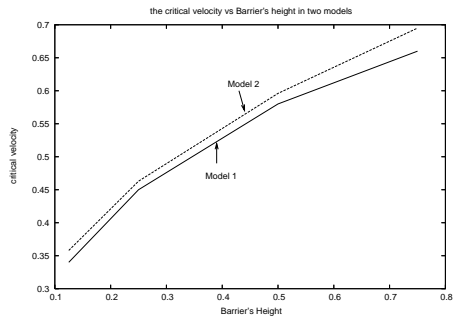}
\caption{Critical velocity as a function of barrier's height in the two models}
\end{center}
\end{figure}

A soliton is an extended object. However, in a  particle picture, the soliton rest mass energy is 
\begin{equation}
\,E=\,M\,c^{2},\quad \mbox{with}\quad c=1
\end{equation}       
where $M$  is the mass of the soliton. Thus, in the particle picture, if a soliton is sent towards a potential barrier with a critical velocity, $u_{cr}$, then it would almost have enough energy to climb up the potential barrier. In this situation the soliton critical energy, $ E_{cr}$,  has to be very close to the soliton rest mass. In a classical particle picture if the kinetic energy of the soliton is completely converted into the potential energy, the soliton will forever stay at the top of the barrier because the difference between the critical energy and the rest mass energy is zero. However, practically, the difference is very small and if we let the difference to be around $\sim x$ where $x$ is a very small number, then 
\begin{center}
$\,E_{cr} - \,M_{rest} \sim \,x,$
\end{center}
where $\,M_{rest}$ is the rest mass of the soliton at the top of the barrier and we expect $x$ to be very small.
\begin{equation}
\,E_{cr}=\frac{\,M}{\sqrt{1-\,u_{cr}^{2}}},
\end{equation}
where $M$ is the rest mass of soliton when it is far away from the potential which was found to be $1.88557$. Thus, the critical velocity is given by
\begin{equation}
\,u_{cr}=\sqrt{1-\left(\frac{\,M}{\,M_{rest}+\,x}\right)^{2}}.
\end{equation}

In model 2 the rest mass of the soliton at the barrier of height 0.5 was found to be 2.3093, see fig 7. Putting $x=0.02$ we estimate the critical velocity as
\begin{center}
$\,u_{cr}=\sqrt{1-\left(\frac{1.88557}{0.02+2.309}\right)^{2}} \sim 0.592,$
\end{center}
which is in excellent agreement with what was found numerically (see fig.8  where it is given by  0.596).

\section{Hole Scattering}

A classical particle is always transmitted through a  potential hole. However, numerical simulations of the scattering of topological solitons [1-2] 
have shown that, when the potential hole is deep enough, there is a critical velocity below which the soliton gets trapped in the hole.
Moreover, just below this critical velocity, the soliton can get reflected by the hole, thus showing
behaviour which is more like of  a quantum object. Of course, there is nothing `quantum' about it - the soliton gets trapped in the hole
and interacts with the radiation waves that it has sent out and ... occasionally, for very specific values of its velocity this interaction
leads to the back escape of the soliton which lookes like its reflection by the hole.

In the present work we have also performed many simulations of the scattering of solitons on the holes for the two models 
and have seen the same behaviour. Like in [1] we have found that the soliton get get trapped in the hole and that it can get reflected.
In all cases it radiates some energy hence, if it comes out of the hole its velocity is lower than at its entry.

In our work we placed a square hole of width 10 at the same position as the barrier
in the previous simulations. Again, when the soliton is far away from the hole its rest mass $M=1.88557$. When the soliton approaches the hole of -0.50 depth, it sees the hole as in figure 9 with no difference between the two models. 
In figure 10  we present a plot of the soliton rest mass as a function of the hole depth.

As, when the soliton is in the hole, it has a lower rest mass in the model 2 than in the model 1, the soliton in model 2
has more `spare' energy and so it it radiates it more.  Figure 11 shows how the position of a soliton moving with a velocity of  0.65 evolves in time in the two models. From the figure we can see that the loss of energy in the  model 2 is higher than in the model 1.

\begin{figure}
\begin{center}
\includegraphics[angle=270, width=6cm]{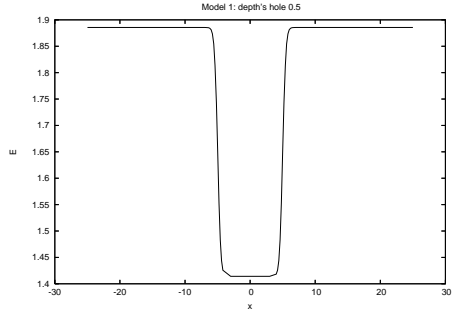}
\caption{The potential hole as seen by the soliton}
\end{center}
\end{figure}



\begin{figure}
\begin{center}
\includegraphics[angle=270, width=6cm]{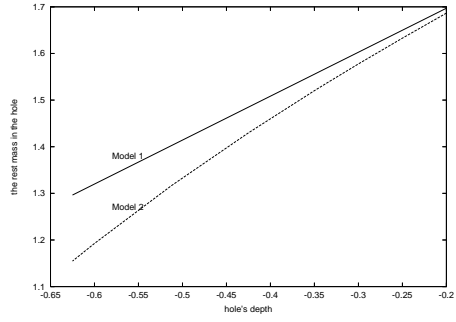}
\caption{Soliton rest mass vs hole's depth in the two models}
\end{center}
\end{figure}

\begin{figure}
\begin{center}
\includegraphics[angle=270, width=6cm]{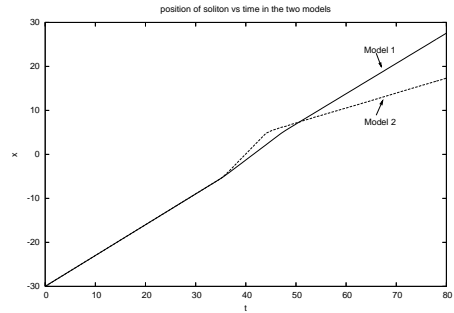}
\caption{Position of the soliton vs time in the two models}
\end{center}
\end{figure}

Incidentally the model 2 has one more interesting scattering property. We have found that for a hole depth just below -0.625, the solitons are always  trapped. Hence there is no critical velocity.  For example at a hole depth of -0.625 the critical velocity is $\sim 0.97$. Figure 12 shows a plot of the critical velocity as a function of depth of  the hole for model 2.

\begin{figure}
\begin{center}
\includegraphics[angle=270, width=6cm]{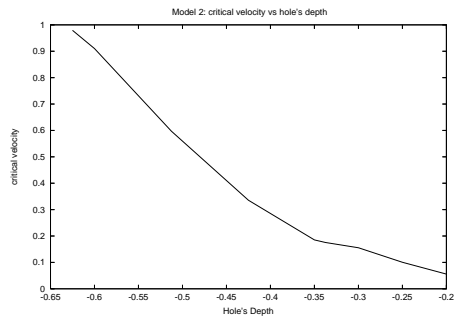}
\caption{Critical velocity vs hole's depth}
\end{center}
\end{figure}

\section{Conclusion}

We have constructed two models involving two different ways of introducing a localised obstruction in the $\lambda\varphi^4$ model and on the scattering 
properties of solitons on obstructions in such models. In the first model the coefficient $\lambda$ was made to be a function
of the space variable $x$,  in the second one the obstruction was introduced through the modification of the metric.
The scattering properties of solitons in both models were very similar, and quite similar to what was seen in the Sine-Gordon model; however
when we looked at the details of the scattering we did spot some differences. The scattering in the first model was more elastic; in each case
we related this to the fact that the solitons had different rest mass energies at the obstruction. The origin of such mass difference in the two models is due to the change of the soliton slope as the soliton climbs the obstruction. Due to this difference the scattering of the soliton
on the obstruction was more elastic in model 1 than in model 2

When we looked at the scattring by the hole we found that, like in the Sine-Gordon case, the scattering was much more inelastic in both models
with model 2 generating much more radiation. In fact, in model 2, for holes sufficiently deep the soliton was completely trapped in them;
{\it ie} we could not find any velocity above which the soliton could escape from the hole.

The question the arises whether all these results are very generic, {\it ie} hold in most (1+1) dimensional solitonic models or are just
the property of the $\lambda \varphi^4$ model. This problem is currently under investigation.
Moreover, we need also to obtain some analytical understanding of the observed results.
In the Sine-Gordon model this was achieved \cite{two} by the introduction of effective variables
describing the oscillation of the vacuum in the hole  - very much in the way Goodman et al \cite{Goodman}
explained the results on Fei et al \cite{older}. We plan to look for a similar explanation 
in this case too.

\end{document}